\documentclass[10pt]{article}
\usepackage{amsmath, amssymb}
\usepackage[utf8]{inputenc}
\usepackage{bm}
\usepackage{breqn}
\usepackage{caption}
\usepackage{graphicx}
\usepackage{setspace}

\title{On the evaluation of uncertainties for state estimation with the Kalman filter}
\author{S. Eichstädt, N. Makarava and C. Elster \\
{\small Physikalisch-Technische Bundesanstalt, Berlin, Germany}}
\date{}

\begin{document}
\begin{spacing}{1}
\maketitle

\vspace{10pt}

\begin{abstract}
The Kalman filter is an established tool for the analysis of dynamic systems with normally distributed noise, and it has been successfully applied in numerous application areas. It provides sequentially calculated estimates of the system states along with a corresponding covariance matrix. For nonlinear systems, the extended Kalman filter is often used which is derived from the Kalman filter by linearization around the current estimate.\\
A key issue in metrology is the evaluation of the uncertainty associated with the Kalman filter state estimates. The ``Guide to the Expression of Uncertainty in Measurements" (GUM) and its supplements serve as the de facto standard for uncertainty evaluation in metrology. We explore the relationship between the covariance matrix produced by the Kalman filter and a GUM-compliant uncertainty analysis. In addition, also the results of a Bayesian analysis are considered. For the case of linear systems with known system matrices, we show that all three approaches are compatible. When the system matrices are not precisely known, however, or when the system is nonlinear, this equivalence breaks down  and different results can be reached then. Though for precisely known nonlinear systems the result of the extended Kalman filter still corresponds to the linearized uncertainty propagation of GUM.\\
The extended Kalman filter can suffer from linearization and convergence errors. These disadvantages  can be avoided to some extent by applying Monte Carlo procedures, and we propose such a method which is GUM-compliant and can also be applied online during the estimation.\\
We illustrate all procedures in terms of a two-dimensional dynamic system and compare the results with those obtained by particle filtering, which has been proposed for the approximate calculation of a Bayesian solution. Finally, we give some recommendations based on our findings.
\end{abstract}

\vspace{2pc}
\noindent{\it Keywords}: Kalman filter, measurement uncertainty, Monte Carlo

\vspace{1pc}

\section{Introduction}
The Kalman filter is a standard approach in signal processing and control theory, and there is a huge number of applications where it has been successfully employed for estimation of system states \cite{Garcia:2013kq, Kollar:1988ek, Heins:2014uv}. Although originally developed for linear systems with normally distributed errors \cite{Kalman, Kalman-Bucy}, several extensions and related methods have been developed since \cite{Julier:2004go}.
The challenge in the application of any signal processing method in metrology is the evaluation of uncertainties in compliance with the "Guide to the Expression of Uncertainty in Measurement" (GUM) \cite{GUM} and its supplements (GUM S1 and GUM S2) \cite{GUMS1, GUMS2}. Here we focus on supplement~2 to the GUM, which addresses multivariate measurands. For the evaluation of uncertainties, GUM~S2 contains a propagation of covariance matrices based on a linearization of the measurement model and a multivariate Monte Carlo method for the propagation of probability density funtions; more details are given in Section \ref{sec:Notation}.  
The Kalman filter calculates estimates and corresponding covariance matrices of the system states sequentially. It is common practice in the signal processing literature to consider these covariance matrices as a measure of confidence or estimation quality. However, a study of their relation to a GUM-compliant uncertainty evaluation is not yet available. 

Dynamic metrology \cite{Eichstadt2012Diss} often is concerned with dynamic systems in which the model is not precisely known. Direct application of the Kalman filter does not account for this, and evaluating the uncertainty associated with obtained state estimates is particularly challenging then.
In the field of signal processing and control theory, the Kalman filter and related approaches for uncertain system models have been addressed by many authors in the field of signal processing and control theory, see, for instance, \cite{LEONDES:2007cz, Souto:2009bj, Xia:2005wb} and references therein. However, the focus of these studies and the aim of the proposed methods is rather a \emph{robustified} state estimation than a reliable evaluation of measurement uncertainties. That is, the uncertainties are considered as a ``defect'' of the measurement setup that has to be mitigated by an adapted estimation method. In addition, many authors consider norm-bounded uncertainties, hence treating uncertainty as a worst-case limit for the obtained estimates \cite{LEONDES:2007cz}. Consequently, these approaches cannot be considered a GUM-compliant treatment of measurement uncertainties.

This paper provides GUM-compliant uncertainty evaluation procedures for the most common scenarios, including linear and nonlinear systems with and without uncertainty in the system model. We discuss and assess the relation of the state error covariance matrix of the Kalman filter to a GUM-compliant measurement uncertainty associated with the state estimates for the different settings. We show that for linear systems with known model, the Kalman filter, the application of GUM Monte Carlo and a Bayesian inference yield compatible results. For linear systems with uncertainty in the model, we consider two approaches; application of GUM Monte Carlo for the linear system and augmentation of the system states to include the uncertain model parameters. In the latter case the original linear dynamic system becomes nonlinear and the standard Kalman filter can no longer be applied. Therefore, we consider the often-used extended Kalman filter which is derived from the Kalman filter by linearization around the current state estimate. For the evaluation of uncertainties, we derive a GUM-compliant measurement model based on the extended Kalman filter. We show that for this model, the estimates and the corresponding covariance matrices calculated by the extended Kalman filter equal those obtained by application of linearized GUM. However, the linearization in the extended Kalman filter can result in significant estimation errors. Therefore, we also propose a GUM Monte Carlo method which, to some extent, is able to mitigate the impact of the linearization onto the state estimate and its corresponding covariance matrix.

The paper is organized as follows. In Section \ref{sec:Notation} we introduce the notation and the assumptions utilized throughout. Section~\ref{sec:LKF} provides uncertainty evaluation for the Kalman filter for linear state-space systems, and Section \ref{sec:EKF} addresses the corresponding methods for the extended Kalman filter in case of nonlinear systems. In Section \ref{sec:MC} we propose a GUM Monte Carlo method for the propagation of uncertainties through the linear Kalman filter and the extended Kalman filter. For general state-space systems, we recommend Bayesian inference and related approximate approaches, and we give a more detailed discussion of their relation to the GUM framework in Section \ref{sec:Particle}. We illustrate the uncertainty propagation by a small  example in Section \ref{sec:example}. Finally, conclusions and recommendations are presented in Section \ref{sec:conclusions}.

\section{Notation}
\label{sec:Notation}
The original Kalman filter \cite{Kalman} has been proposed for state estimation in linear systems
\begin{align}
\bm{x}(k+1) =& \bm{F}(k)\bm{x}(k) + \bm{D}(k)\bm{w}(k) \label{linear_state}\\
\bm{y}(k) =& \bm{C}(k)\bm{x}(k) + \bm{v}(k) \label{linear_obs}
\end{align}
with possibly time-varying system model matrices $\bm{F, C, D}$, state vector $\bm{x}(k)$, measurement vector $\bm{y}(k)$ and independent Gaussian noise processes $\bm{w}(k)$ and $\bm{v}(k)$ with known covariance matrices $\bm{Q}(k)$ and $\bm{R}(k)$, respectively. For ease of presentation we consider $\bm{D}\equiv \bm{I}$, i.e., white noise in the state equation. The conclusions of our study remain basically the same for other choices of $\bm{D}$. Throughout, we use canonical notation from system theory. That is, capital bold variables represent matrices, lower case bold variables represent vectors, and an estimate of a quantity $\bm{x}$ is denoted by $\hat{\bm{x}}$.

The aim in the here considered state estimation is the sequential estimation of $\bm{x}(k)$ from knowledge about the system model matrices and the observations $\bm{y}(k)$. In a sequential estimation an estimate $\hat{\bm{x}}(k)$ with associated covariance matrix $\bm{P}(k)$ is calculated from the estimate and its covariance matrix at the previous time step $k-1$ and the measurement $\bm{y}(k)$ at the current time step, starting with an initial estimate and an initial covariance matrix for time $k=0$. The corresponding Kalman filter equations are given by \cite{Kalman}
\begin{align}
	\bm{x}_{k,k-1} =& \bm{F}(k)\hat{\bm{x}}(k-1) \label{LK:predict_state}\\
	\bm{P}_{k,k-1} =& \bm{F}(k)\hat{\bm{P}}(k-1)\bm{F}^{\mathsf{T}}(k)+\bm{Q}(k) \label{LK:predict_cov}\\
	\hat{\bm{x}}(k) =& \bm{x}_{k,k-1} + \bm{K}(k) \left( \hat{\bm{y}}(k) - \bm{C}(k)\bm{x}_{k,k-1} \right) \label{LK:est_state} \\
	\hat{\bm{P}}(k) =& \left( \bm{I} - \bm{K}(k)\bm{C}(k)\right)\bm{P}_{k,k-1} \label{LK:est_cov}
\end{align}
with Kalman gain matrix
\begin{equation}
	\bm{K}(k) = \bm{P}_{k,k-1}\bm{C}^{\mathsf{T}}(k) \left( \bm{C}(k)\bm{P}_{k,k-1}\bm{C}^{\mathsf{T}}(k) + \bm{R}(k) \right)^{-1} .
\label{LK:gain}
\end{equation}
Equations (\ref{LK:predict_state})-(\ref{LK:predict_cov}) denote the so-called prediction, and equations (\ref{LK:est_state})-(\ref{LK:est_cov}) the correction step. The Kalman filter equations can be derived in various ways, e.g. using orthogonal projection \cite{Kalman} or Bayes' theorem \cite{Candy}, see also Section \ref{sec:LKF}. In the classical interpretation, the matrix $\bm{P}(k)$ represents the error covariance matrix $\mathbb{E} \left[ (\hat{\bm{x}}(k)-\bm{x}(k))^2 \right]$, whereas in a Bayesian interpretation $\bm{P}(k)$ is the covariance matrix of the posterior distribution $\pi(\bm{x}(k)\vert \bm{x}(k-1),\bm{y}(k))$ \cite{Candy}. 
The Kalman filter propagates the initial covariance matrix $\bm{P}(0)$ sequentially from time instant $k=0$ using information about the state covariance matrix $\bm{Q}$, the measurement covariance matrix $\bm{R}$ and the Kalman gain $\bm{K}$. 

One aim of this paper is to address the relation of the matrix $\bm{P}(k)$ to a GUM-compliant uncertainty expression for the estimate $\hat{\bm{x}}(k)$. Therefore, we consider GUM supplement~2 (GUM~S2), which addresses multivariate quantities \cite{GUMS2}. For the evaluation of uncertainties, GUM~S2 contains two approaches: a propagation of estimates and covariance matrices based on a linearization of the measurement model, and a propagation of probability density functions (PDFs) using a Monte Carlo method. The application of the GUM framework for the evaluation of uncertainties requires a precise definition of the measurand, the mathematical model for its calculation, and estimates of the input quantities together with a corresponding covariance matrix (linearized GUM) or a PDF (GUM Monte Carlo) associated with the model input quantities. For the linearized GUM method, estimates and their associated covariance matrix for  the input quantities are propagated through the linearized measurement model. The GUM Monte Carlo method does not require a linearization, but instead propagates samples drawn from the (joint) PDF of the input quantities through the measurement model. The result is a set of samples of the model output, which are considered a discrete representation of the PDF associated with the measurand. The estimate of the measurand and its associated uncertainty are then obtained as the mean and covariance matrix of that PDF, respectively.

For time-dependent measurands, two approaches are possible. When a closed time interval is considered, the whole sequence $\bm{x}= \left( \bm{x}(1),\ldots,\bm{x}(N) \right)^{\mathsf{T}}$ can be taken as measurand. We refer to this as \emph{batch estimation}. For a sequence with one-dimensional elements $x(k)$, the result of batch estimation is an estimate $\hat{\bm{x}} = \left( \hat{x}(0),\ldots,\hat{x}(N) \right)^{\mathsf{T}}$ with an associated covariance matrix for linearized GUM or, respectively, a multivariate PDF for GUM Monte Carlo. In the case of multivariate elements $\bm{x}(k)$, either point-wise estimates and covariance matrices at the individual time instants are considered, or a re-ordering of the sequence is required.
Batch estimation with the GUM Monte Carlo method is clearly challenging in terms of required computer resources for the case of longer time intervals. Therefore, one can consider as measurand the individual $\bm{x}(k)$ instead of the whole sequence $\bm{x}$. We refer to this as \emph{sequential estimation}. In this way, the dimension of the Monte Carlo draws does not increase with the length of the time interval, and required computer resources can be significantly reduced, see also \cite{Eichstadt2012, Eichstadt2012Diss}. The result of sequential estimation of the measurand $\bm{x}(k)$ is an estimate $\hat{\bm{x}}(k)$ with associated covariance matrix $\bm{U}_{\bm{x}(k)}$ for the linearized GUM approach, and a multivariate PDF associated with $\bm{x}(k)$ for the GUM Monte Carlo method.

\section{Uncertainty propagation for the linear Kalman filter}
\label{sec:LKF}
Uncertainty propagation in line with the GUM requires a measurement model for the measurand together with estimates and associated uncertainties for the input quantities (GUM), or a joint probability density function (PDF) as a state-of-knowledge distribution associated with the input quantities (GUM S1 and GUM S2). For the application of the linear Kalman filter (\ref{LK:predict_state})-(\ref{LK:gain}) the measurement model is given by
\begin{align}
	\bm{x}(k) =& \bm{x}_{k,k-1} + \bm{K}(k)(\bm{y}(k)-\bm{C}(k)\bm{x}_{k,k-1}) \\
	          =& (\bm{I} - \bm{K}(k)\bm{C}(k))\bm{x}_{k,k-1} + \bm{K}(k)\bm{y}(k) \label{eq:LKF_Xmodel}
\end{align}
where
\[ \bm{x}_{k,k-1} = \bm{F}(k)\bm{x}(k-1) + \bm{z}(k) \]
with $\bm{z}(k)\sim N(0,\bm{Q}(k))$. Basically two scenarios have to be addressed: known measurement model and uncertain measurement model.

\subsection{Known measurement model}
\label{sec:LKF_knownmodel}
When the system model is considered to be known exactly, the only sources of uncertainty 
in the prediction of the current state estimate in (\ref{eq:LKF_Xmodel}) are the measurement $\bm{y}(k)$ and the state noise $\bm{z}(k)$. The uncertainty associated with $\bm{y}(k)$ is given by the covariance matrix $\bm{R}(k)$, which represents the noise covariance in the observations at time instant $k$. In the simplest case, the noise is assumed to be identically, independently distributed (i.i.d.) zero-mean Gaussian noise (i.e. white) with known standard deviation. For more complex noise structures, vector auto-regressive (VAR) models can be employed \cite{Reinsel:1992ki}. 

Assume that $\bm{P}(k-1)=\bm{P}_{k-1,k-1}$ associated with $\hat{\bm{x}}(k-1)$ represents a valid GUM-compliant uncertainty. Then, according to GUM Supplement~2 \cite{GUMS2}, the multivariate normal distribution 
\[\bm{x}_{k,k-1} \sim\mathcal{N} \left( \bm{F}(k)\hat{\bm{x}}(k-1), \bm{F}(k)\bm{P}_{k-1,k-1}\bm{F}^{\mathsf{T}}(k) + \bm{Q}(k) \right)\]
models the state-of-knowledge about $\bm{x}_{k,k-1}$. 
Note that $\bm{Q}(k)$ is the covariance matrix associated with the employed estimate $\bm{0}$ for $\bm{z}(k)$.

Knowledge about the measurement $\bm{y}(k)$ is modeled by the PDF
\[ \bm{y}(k) \sim \mathcal{N} \left( \hat{\bm{y}}(k),  \bm{R}(k) \right) .\]
Both input quantities are assumed to be obtained independently, which yields their joint distribution as the product of their associated PDFs. Since the measurement model (\ref{eq:LKF_Xmodel}) is linear in $\bm{x}_{k,k-1}$ and $\bm{y}(k)$, and as the state-of-knowledge distributions are normal distributions, the propagation of the joint PDF through the measurement model can be carried out analytically, resulting in the normal distribution with mean
\begin{equation}
	\hat{\bm{x}}(k) = \left( \bm{I} - \bm{K}(k)\bm{C}(k) \right)\bm{F}(k)\hat{\bm{x}}(k-1) + \bm{K}(k)\hat{\bm{y}}(k)
\end{equation}
and covariance matrix
\begin{align}
\hspace{-8ex}		\bm{U}_{\bm{x}(k)} =&\left( \bm{I}-\bm{K}(k)\bm{C}(k) \right)\bm{P}_{k,k-1}(\bm{I}-\bm{K}(k)\bm{C}(k))^{\mathsf{T}} + \bm{K}(k)\bm{R}(k)\bm{K}(k)^{\mathsf{T}} \\
\hspace{-8ex}	        =& \bm{P}_{k,k-1} - \bm{P}_{k,k-1} \bm{C}(k)^{\mathsf{T}}\bm{K}(k)^{\mathsf{T}} - \bm{K}(k)\bm{C}(k)\bm{P}_{k,k-1} + \\
\hspace{-8ex}	        &\bm{K}(k)\bm{C}(k)\bm{P}_{k,k-1}\bm{C}(k)^{\mathsf{T}}\bm{K}(k)^{\mathsf{T}} + \bm{K}(k)\bm{R}(k)\bm{K}(k)^{\mathsf{T}} \\
\hspace{-8ex}	        =&\left( \bm{I}-\bm{K}(k)\bm{C}(k) \right)\bm{P}_{k,k-1} ,
\end{align}
where $\bm{P}_{k,k-1} = \bm{F}(k)\bm{P}_{k-1,k-1}\bm{F}^{\mathsf{T}}(k) + \bm{Q}(k)$.
Hence, the PDF associated with the measurand $\bm{x}(k)$, calculated by a propagation of the joint PDF associated with the input quantities $\bm{y}(k)$ and $\bm{x}_{k,k-1}$ through the Kalman filter estimation equation (\ref{eq:LKF_Xmodel}) corresponds to the outcome of the linear Kalman filter itself. That is, the Kalman filter state estimate $\hat{\bm{x}}(k)$ can be considered as an estimate of the measurand and the Kalman filter covariance matrix estimate $\bm{P}(k)=\bm{P}_{k,k}$ equals the uncertainty associated with the measurand in the sense of the GUM.

It is well known that the linear Kalman filter for Gaussian independent noise processes $\bm{w}_k$, $\bm{v}_k$ with known covariance matrices $\bm{Q}(k)$ and $\bm{R}(k)$ corresponds to a Bayesian inference \cite{Candy} with 
\begin{itemize}
 \item prior
 \begin{equation}
\hspace{-8ex}	\bm{x}(k) \vert \bm{x}(k-1) \sim N\left( \bm{F}(k)\hat{\bm{x}}(k-1), \bm{F}(k)\bm{P}(k-1)\bm{F}(k)^{\mathsf{T}}+\bm{Q}(k) \right)	
 \end{equation}
 \item likelihood
 \begin{equation}
\hspace{-18ex}	\bm{y}(k) \vert \bm{x}(k) \sim N\left( \bm{C}(k) \bm{x}(k), \bm{R}(k) \right)
 \end{equation}
 \item posterior
  \begin{align}
  	\bm{x}(k) \vert \bm{y}(k) \sim	& N\left( \bm{x}_{k,k-1}+\bm{K}(k)(\bm{y}(k)-\bm{C}(k)\bm{x}_{k,k-1}), \right. \nonumber\\ 
  			&\left. \left( \bm{I} - \bm{K}(k)\bm{C}(k) \right)\bm{P}_{k,k-1} \right) .
  \end{align}
\end{itemize}

Consequently, the GUM Monte Carlo result under these assumptions produces the same distribution as a Bayesian inference.

\subsection{Uncertainty in the model}
\label{sec:LKF_uncertain}
We here assume that the state-space system matrices depend on a parameter vector $\bm{\theta}=(\theta_1,\ldots,\theta_M)^{\mathsf{T}}$ which is not precisely known, i.e. $\bm{F}=\bm{F}(\bm{\theta},k)$ and $\bm{C}=\bm{C}(\bm{\theta},k)$. The corresponding equations for the correction step of the Kalman filter are then given by
\begin{align}
	\bm{x}(k) =& \bm{x}_{k,k-1} + \bm{K}(\bm{\theta},k) \left( \bm{y}(k) - \bm{C}(\bm{\theta},k)\bm{x}_{k,k-1} \right) \\
	\bm{P}(k) =& \bm{P}_{k,k-1} - \bm{K}(\bm{\theta},k) \left( \bm{C}(\bm{\theta},k)\bm{P}_{k,k-1}\bm{C}^{\mathsf{T}}(\bm{\theta},k) + \bm{R}(k) \right)
\end{align}
with Kalman gain
\begin{equation}
	\hspace{-8ex}\bm{K}(\bm{\theta},k) = \bm{P}_{k,k-1}\bm{C}^{\mathsf{T}}(\bm{\theta},k) \left( \bm{C}(\bm{\theta},k)\bm{P}_{k,k-1}\bm{C}^{\mathsf{T}}(\bm{\theta},k)+\bm{R}(k) \right)^{-1} .
\end{equation}
The measurement model for the measurand $\bm{x}(k)$ is thus given by
\begin{align}
	\hspace{-8ex}\bm{x}(k) =& \bm{F}(\bm{\theta},k)\bm{x}(k-1) + \bm{K}(\bm{\theta},k) \left( \bm{y}(k) - \bm{C}(\bm{\theta},k)\bm{F}(\bm{\theta},k)\bm{x}(k-1) \right) \\
	 =& \bm{x}_{k,k-1} + \bm{K}(\bm{\theta},k)\left( \bm{y}(k) - \bm{C}(\bm{\theta},k)\bm{x}_{k,k-1} \right) \label{eq:LKF_GUMmodel}
\end{align}
with input quantities now being $\bm{x}_{k,k-1}$, $\bm{y}(k)$ and $\bm{\theta}$. In contrast to the case considered in the previous section, the measurement model is nonlinear in the input quantities. In principle, a linearization using, for instance, finite differences could be applied in line with GUM~\cite{GUM}. However, owing to the nonlinearity of the model, significant estimation errors can result, and a Monte Carlo method in accordance with GUM Supplement~2 \cite{GUMS2} is recommended instead, see also Section \ref{sec:MC}. It is worth noting that a Bayesian approach to state-estimation as in the previous section does in general not provide a closed formula solution due to the nonlinearity of the model. Therefore, a Markov chain Monte Carlo method or similar techniques would have to be employed in order to calculate the posterior PDF of the state estimate $\bm{x}(k)$, cf. Section~\ref{sec:Particle}.

An often applied technique for the treatment of the model parameter $\bm{\theta}$ is the augmentation of the state-space model (\ref{linear_state})- (\ref{linear_obs}) such that the state vector also contains the parameter vector $\bm{\theta}$:
\begin{equation}
\left( \begin{array}{c}
\bm{x}(k+1) \\ \bm{\theta}(k+1)
\end{array}\right)
=
\left( \begin{array}{c}
F(\bm{\theta},k)\bm{x}(k) \\ \bm{\theta}(k)
\end{array}\right)
+
\left( \begin{array}{c}
\bm{w}(k) \\ 0
\end{array}\right)
\end{equation}
with initial state error covariance matrix then defined as
\begin{equation}
	\bm{P}(0) = \left(\begin{array}{cc}
\bm{P}^{x}(0) & 0 \\ 0 & \bm{U_{\bm{\theta}}}
\end{array}\right) .
\end{equation}
The resulting model is a nonlinear state-space model and the linear Kalman filter can no longer be applied. However, many modifications of the linear Kalman filter for nonlinear models can be found in the literature, the most famous being the so called extended Kalman filter, see also Section~\ref{sec:EKF}. One outcome of such a modified Kalman filter is the state estimate error covariance matrix $\bm{P}(k)$ which for the above model can be written in the following way
\begin{equation}
\bm{P}(k) = \left(\begin{array}{cc}
\bm{P}^{\bm{x}}(k) & \bm{P}^{\bm{x},\bm{\theta}}(k) \\ (\bm{P}^{\bm{x},\bm{\theta}}(k))^{\mathsf{T}} & \bm{P}^{\bm{\theta}}(k)
\end{array}\right) .
\end{equation}
The question arises whether the sub-matrix $\bm{P}^{\bm{x}}(k)$ can be considered a GUM-compliant representation of the uncertainty associated with $\hat{\bm{x}}(k)$. In the following we show that this is not the case when the extended Kalman filter is applied for state estimation.

The prediction and correction step of the Kalman filter and related approaches can be considered in separate equations for $\bm{x}(k)$ and $\bm{\theta}(k)$~\cite{Ljung:1979vz}
\begin{align}
	\hat{\bm{x}}(k) =& \bm{x}_{k,k-1} + \bm{K}_1(k)\left( \bm{y}(k)-\bm{C}(\bm{\theta}_{k,k-1},k)\bm{x}_{k,k-1} \right) \\
	\hat{\bm{\theta}}(k) =& \bm{\theta}_{k,k-1} + \bm{K}_2(k) \left( \bm{y}(k)-\bm{C}(\bm{\theta}_{k,k-1},k)\bm{x}_{k,k-1} \right) \label{eq:update_theta}\\
	\bm{P}^x(k) =& \bm{P}_{k,k-1}^{x} - \bm{K}_1(k)\left( \bm{H}_{\bm{\theta}}\bm{P}_{k,k-1}\bm{H}^{\mathsf{T}}_{\bm{\theta}}+\bm{R}(k) \right)\bm{K}_1^{\mathsf{T}}(k) \\
	\bm{P}^{\bm{\theta}}(k) =& \bm{P}_{k,k-1}^{\bm{\theta}} - \bm{K}_2(k)\left( \bm{H}_{\bm{\theta}}\bm{P}_{k,k-1}\bm{H}^{\mathsf{T}}_{\bm{\theta}}+\bm{R}(k) \right)\bm{K}_2^{\mathsf{T}}(k) \label{eq:update_Ptheta} \\
	\bm{P}^{\bm{x,\theta}}(k) =& \bm{P}_{k,k-1}^{\bm{x,\theta}} - \bm{K}_1(k)\left( \bm{H}_{\bm{\theta}}\bm{P}_{k,k-1}\bm{H}^{\mathsf{T}}_{\bm{\theta}}+\bm{R}(k) \right)\bm{K}_2^{\mathsf{T}}(k)
\end{align} 
with
\begin{align*} 
\bm{H}_{\bm{\theta}} =& \left.\left( \bm{C}(\bm{\theta},k),\frac{\mathrm{d}}{\mathrm{d}\bm{\theta}} \bm{C}(\bm{\theta},k)\bm{x} \right)\right|_{\bm{\theta}_{k,k-1},\bm{x}_{k,k-1}} \\
					=:& \left( \bm{C}(\bm{\theta}_{k,k-1},k), \bm{D}(\bm{\theta}_{k,k-1},\bm{x}_{k,k,-1} \right)
\end{align*}

and
\[
\hspace{-8ex}
	\bm{K}_1(k) = \left( \bm{P}_{k,k-1}^{x}\bm{C}(\bm{\theta}_{k,k-1},k) + \bm{P}_{k,k-1}^{\bm{\theta,x}}\bm{D}(\bm{\theta}_{k,k-1},\bm{x}_{k,k-1}) \right)\bm{S}(\bm{\theta},k)^{-1}
\]
\[
\hspace{-8ex}
	\bm{K}_2(k) = \left( \bm{P}_{k,k-1}^{\bm{\theta}}\bm{C}(\bm{\theta}_{k,k-1},k) + \bm{P}_{k,k-1}^{\bm{\theta,x}}\bm{D}(\bm{\theta}_{k,k-1},\bm{x}_{k,k-1}) \right)\bm{S}(\bm{\theta},k)^{-1}
\]
where 
\[ \bm{S}(\bm{\theta},k) =  \bm{H}_{\bm{\theta}}\bm{P}_{k,k-1}\bm{H}^{\mathsf{T}}_{\bm{\theta}}+\bm{R}(k) .\]

Both parts of the state update depend on the parameter vector $\bm{\theta}$. In particular, during the Kalman filter process, equation (\ref{eq:update_theta}) updates the estimate of $\bm{\theta}$ and equation (\ref{eq:update_Ptheta}) its associated uncertainty. Hence, the extended Kalman filter and any related approach do not solely propagate the uncertainty associated with the model parameters, but also alter their estimated value. This shows that with the extended Kalman filter for the augmented model, both, the original state vector $\bm{x}(k)$ and the model parameter vector $\bm{\theta}$, are estimated. With regard to the original measurement model (\ref{eq:LKF_GUMmodel}), this would make $\bm{\theta}$ an input quantity as well as an output quantity of the measurement model, rendering the model non-compliant with GUM. On the other hand, a Bayesian inference as for the linear case can be carried out for the case of uncertain systems in a natural way. That is, one may calculate the joint posterior of $\bm{x}(k)$ and $\bm{\theta}$ at time instant $k$ and obtain the posterior of $\bm{x}(k)$ by marginalization, treating $\bm{\theta}$ as a nuisance parameter. In this way, the estimate of the parameter $\bm{\theta}$ changes with time as new measurements $\bm{y}(k)$ are taken into account.

A GUM-compliant measurement model for the case of uncertainty in the model can be obtained by treating the parameter $\bm{\theta}$ as time-dependent. That is, $\bm{\theta}$ and $\bm{U}_{\bm{\theta}}$ are interpreted as initial values for the extended Kalman filter just like $\bm{x}(0)$ and $\bm{P}(0)$. The parameters $\bm{\theta}(k)$ at time instants $k>0$ are then considered as individual measurands, i.e., one considers $\bm{\Theta}=\left( \bm{\theta}(1),\bm{\theta}(2),\ldots,\bm{\theta}(N) \right)^{\mathsf{T}}$. As a result, at time instant $k$ the parameter $\bm{\theta}(k-1)$ is the input quantity and $\bm{\theta}(k)$ the output quantity of the measurement model, which is thus in accordance with GUM. Application of the extended Kalman filter then produces an estimate of the augmented state vector and its covariance matrix. In the following section we show that the result of the extended Kalman filter is equivalent to that obtained by an application of linearized GUM.

\section{Uncertainty propagation for the extended Kalman filter}
\label{sec:EKF}

The extended Kalman filter (EKF) aims at state estimation for nonlinear state-space models of the form
\begin{align}
	\bm{x}(k+1) =& f(\bm{x}(k),k) + \bm{w}(k) \label{ekf_state}\\
	\bm{y}(k) =& h(\bm{x}(k),k) + \bm{v}(k) \label{ekf_obs}
\end{align}
by linearizing around the current estimate $\hat{\bm{x}}(k)$ and applying the Kalman filter \cite{Julier:2004go}. The corresponding equations for the model (\ref{ekf_state})-(\ref{ekf_obs}) are given by
\begin{align}
	\bm{x}_{k,k-1} =& f(\hat{\bm{x}}(k-1),k) \label{EKF:pred_state}\\
	\bm{P}_{k,k-1} =& \bm{F}(k)\hat{\bm{P}}(k-1)\bm{F}^{\mathsf{T}}(k)+\bm{Q}(k) \label{EKF:pred_cov}\\
	\hat{\bm{x}}(k) =& \bm{x}_{k,k-1} + \bm{K}(k)\left( \bm{y}(k)-h(\bm{x}_{k,k-1},k) \right) \label{EKF:est_state} \\
	\hat{\bm{P}}(k) =& \left( \bm{I}-\bm{K}(k)\bm{H}(k) \right)\bm{P}_{k,k-1} \label{EKF:est_cov}
\end{align}
with Kalman gain $\bm{K}(k)$ calculated as
\[ \bm{K}(k) = \bm{P}_{k,k-1}\bm{H}^{\mathsf{T}}(k)\left( \bm{H}(k)\bm{P}_{k,k-1}\bm{H}^{\mathsf{T}}(k)+\bm{R}(k) \right)^{-1} \]
and derivatives
\[ \bm{F}(k)=\left.\frac{\mathrm{d}}{\mathrm{d}\bm{x}}f(\bm{x}(k),k)\right|_{\bm{x}_{k,k-1}} \textnormal{\quad and \quad} 
   \bm{H}(k)=\left.\frac{\mathrm{d}}{\mathrm{d}\bm{x}}h(\bm{x}(k),k)\right|_{\bm{x}_{k,k-1}} .\]
In contrast to the linear Kalman filter, the extended Kalman filter is known to be a sub-optimal filter \cite{Arulampalam:2002hg} and the amount of the difference to the result of a Bayesian inference depends on the effect of the nonlinearity, i.e. the linearization error of the filter equations.

\subsection{Known measurement model}
Similarly as for the linear state-space model in the previous section, our analysis starts with the uncertainty propagation for the case that at time instant $k$ only the measurements $\bm{y}(k)$, the state noise $\bm{z}(k)$ and the previous state $\bm{x}(k-1)$ are considered as uncertain input quantities for the measurand $\bm{x}(k)$. The measurement model is thus given by
\begin{equation}
	\bm{x}(k) = \bm{x}_{k,k-1} + \bm{K}(k)\left( \bm{y}(k)-h(\bm{x}_{k,k-1},k) \right)  \label{ekf_measurand}
\end{equation}
with input quantities $\bm{x}_{k,k-1}=f(\bm{x}(k-1),k) + \bm{z}(k)$ and $\bm{y}(k)$ as in Section \ref{sec:LKF_knownmodel}. 
In contrast to Section \ref{sec:LKF_knownmodel}, however, the measurement model depends nonlinearly on the input quantity $\bm{x}_{k,k-1}$. 

The Kalman gain $\bm{K}(k)$ is evaluated at the estimate $f(\hat{\bm{x}}(k-1),k)$. The application of the GUM linearization approach then results in
\begin{equation}
\hspace{-10ex} \hat{\bm{x}}(k) = f(\hat{\bm{x}}(k-1),k) + \bm{K}(k)\left( \hat{\bm{y}}(k)-h(f(\hat{\bm{x}}(k-1),k),k) \right)
\end{equation}
with an associated uncertainty calculated as
\begin{align}
 \bm{U}_{\bm{x}(k)} =& \left( \bm{I}-\bm{K}(k)\bm{H}(k) \right)\bm{P}_{k,k-1}\left( \bm{I}-\bm{K}(k)\bm{H}(k) \right)^{\mathsf{T}} + \bm{K}(k)\bm{R}(k)\bm{K}^{\mathsf{T}}(k) \\
		=& \left( \bm{I}-\bm{K}(k)\bm{H}(k) \right)\bm{P}_{k,k-1} .
\end{align}
Hence, the result of GUM is identical to that of the extended Kalman filter. Similar to the difference between extended Kalman filter and a Bayesian inference, the deviation of the extended Kalman filter and a GUM Monte Carlo (GUM S1 or GUM S2) result depends on the effect of the nonlinearity of $f(\cdot)$ and $h(\cdot)$. An efficient sequential implementation of the GUM Monte Carlo method for the extended Kalman filter is given in Section \ref{sec:MC}. Note that the Monte Carlo method does not mitigate the linearization error of the extended Kalman filter itself. That is, the reliability of the estimation still depends on that linearization error. In principle, one could this take into account as an additional uncertainty contribution. However, since the linearization error depends nonlinearly on the estimation error, this is rather difficult in practice. As an alternative, other state estimation methods may have to be considered then, cf. Section \ref{sec:Particle}.

\subsection{Uncertainty in the model}
Similar to the linear Kalman filter in section \ref{sec:LKF}, we assume that the model functions depend on a parameter vector $\bm{\theta}$. Knowledge about $\bm{\theta}$ is assumed to be available in terms of an estimate $\hat{\bm{\theta}}$ and an associated covariance matrix $\bm{U_\theta}$. The aim is to propagate the uncertain knowledge about the model through the application of the extended Kalman filter. As for the case of a linear state-space model, we commence by extending the measurement model for the evaluation of the measurand through $\bm{\theta}$ according to
\begin{equation}
	\bm{x}(k) = \bm{x}_{k,k-1} + \bm{K}(\bm{x}(k-1),\bm{\theta})\left( \bm{y}(k) - h(\bm{x}(k-1),\bm{\theta},k) \right) .
	\label{eq:EKF_model2}
\end{equation} 
It is to be expected that in this way the nonlinearity of the measurement model increases, in the sense that errors owing to a linearization for the propagation of uncertainties increase. Thus, the application of a Monte Carlo method is recommended instead. 

As for the linear model, the parameter $\bm{\theta}$ can be considered for estimation and the state-space model augmented to 
\begin{align}
	\left( \begin{array}{c}
\bm{x}(k+1) \\ \bm{\theta}(k+1)
\end{array}\right)
=
\left( \begin{array}{c}
f(\bm{x}(k),\bm{\theta},k) \\ \bm{\theta}(k)
\end{array}\right)
+
\left( \begin{array}{c}
\bm{w}(k) \\ 0
\end{array}\right)
\end{align}
with initial state error covariance matrix for the extended Kalman filter then defined as
\begin{equation}
	\bm{P}(0) = \left(\begin{array}{cc}
\bm{P}^{x}(0) & 0 \\ 0 & \bm{U_{\bm{\theta}}}
\end{array}\right) .
\end{equation}
Application of the extended Kalman filter then utilizes the covariance matrix $\bm{U_{\bm{\theta}}}$ as initial estimate of the state variance for that parameter and updates the estimate of $\bm{\theta}$ over time.  
With the exception of the definition of $\bm{D}\left( \bm{x}_{k,k-1},\bm{\theta}_{k,k-1} \right)$, the resulting formulas for the extended Kalman filter are the same as for the augmented linear state-space model. Consequently, the same conclusions hold for the nonlinear case. That is, the parameter vector $\bm{\theta}$ and its associated uncertainty $\bm{U}_{\bm{\theta}}$ can be considered as initial values for the extended Kalman filter with augmented state vector. On the other hand, when the parameter $\bm{\theta}$ is not considered as input quantity to the measurement model, the estimated matrix $\bm{P}(k)$ coincides with a linearized GUM uncertainty evaluation, but neither with a Bayesian inference nor with GUM Monte Carlo.

\section{GUM Monte Carlo for Kalman filter methods}
\label{sec:MC}
Supplement~2 to the GUM \cite{GUMS2} describes the propagation of uncertainties for multivariate quantities. Measurement uncertainty in that context is defined as the covariance matrix of a state of knowledge multivariate PDF. Consequently, the estimate is defined as the expectation of that PDF. Propagation of uncertainty then becomes a propagation of the (joint) multivariate PDF associated with the input quantities  by drawing samples from that PDF and evaluating the measurement model for the drawn samples. The result is a sample from the PDF associated with the measurand. The propagation of PDFs via Monte Carlo is related to the change-of-variables formula for the PDF associated with the input quantities~\cite{possolo2007assessment}. 

For sequential measurement models, such as filtering, basically two approaches to the implementation of the Monte Carlo method for uncertainty evaluation are possible: batch Monte Carlo and sequential Monte Carlo. Their main difference is in the definition of the measurand and the resulting dimensionality of the evaluated uncertainty. That is, for batch Monte Carlo the measurand is considered to be the sequence $\bm{X}=\left( \bm{x}(1),\ldots,\bm{x}(N) \right)^{\mathsf{T}}$, whereas for sequential Monte Carlo the measurand is $\bm{x}(k)$ for a particular time instant $k$. In this section we outline batch Monte Carlo and sequential Monte Carlo for the propagation of uncertainty through a Kalman filter method. Both Monte Carlo methods are valid implementations of GUM Monte Carlo and yield identical results. The sequential GUM Monte Carlo, however, can be advantageous when online estimation is considered.

\subsection{Batch Monte Carlo for GUM-compliant uncertainty evaluation}
\label{sec:batchMC}
The Monte Carlo method of GUM Supplement~2 \cite{GUMS2} could, in principle, be applied directly to the linear and nonlinear Kalman filter state estimation, irrespective of its nonlinearity. That is, the measurement model for the estimation of the measurand $\bm{X}=\left( \bm{x}(1),\ldots,\bm{x}(N) \right)^{\mathsf{T}}$ could be considered to be given in terms of the pseudo-code
\begin{verbatim}
DEF ESTIMATE_X(y,theta)
   FOR k in [1,2,...,N]
       x_(k,k-1),P_(k,k-1) = Kalman_predict(x[k-1],theta)
       xtilde = x_(k,k-1) + MULTIVARIATE_NORMAL( 0 , Q[k] )
       x[k] = Kalman_correct(xtilde,P_(k,k-1),y[k],theta)
   ENDFOR
   RETURN x
ENDDEF
\end{verbatim}
where the forecasting and prediction steps depend on the chosen Kalman filter method. Note the intermediate calculation of \verb!xtilde!, which is necessary to account for the uncertainty contribution of the state covariance, cf. Section \ref{sec:LKF}. 

The outcome of the Monte Carlo method is a sample of high-dimensional vectors, with dimension depending on the number of states and the number of time instants considered. The drawback of this approach is thus the amount of required computer memory, which increases drastically for an increased number of time samples. However, the implementation of the batch Monte Carlo method is straightforward and does not require a significant adaptation of the state estimation routine. 

\subsection{Sequential Monte Carlo for GUM-compliant uncertainty evaluation}
For the sequential Monte Carlo method, $\bm{x}(k)$ is considered as measurand and $\bm{x}_{k,k-1}, \bm{y}(k), \bm{\theta}, \bm{z}(k)$ as input quantities, for which knowledge is assumed to be available in terms of the joint PDF 
\begin{equation}
	p_{\bm{y}(k),\bm{z}(k),\bm{x}_{k,k-1},\bm{\theta}}(\bm{\rho}) = p_{\bm{y}(k)}(\bm{\psi})
	p_{\bm{z}(k)}(\bm{\zeta})
	p_{\bm{x}_{k,k-1}, \bm{\theta}}(\bm{\xi}) .
\end{equation}
Following the assumptions made, $p_{\bm{y}(k)}$ and $p_{\bm{z}(k)}$ are both taken as multivariate Gaussian distributions with means $\hat{\bm{y}}(k)$ and $\bm{0}$ together with the covariance matrices $\bm{R}(k)$ and $\bm{Q}(k)$, respectively.
The PDF $p_{\bm{x}_{k,k-1}, \bm{\theta}}$ does not factorize since $\bm{x}_{k,k-1}$ and $\bm{\theta}$
are not independent. This PDF evolves from the PDF $p_{\bm{x}(0), \bm{\theta}}=p_{\bm{x}(0)} p_{\bm{\theta}}$ according to the sequential estimation procedure, where $p_{\bm{x}(0)}$ and $p_{\bm{\theta}}$ are assigned according to the prior knowledge about $\bm{x}(0)$ and $\bm{\theta}$.

Consider the measurement model to be given by the prediction and correction steps of the Kalman filter, i.e. by the following pseudo-code.
\begin{verbatim}
DEF ESTIMATE_Xk(y[k],x[k-1],theta)
   x_(k,k-1),P_(k,k-1) = Kalman_predict(x[k-1],theta)
   xtilde = x_(k,k-1) + MULTIVARIATE_NORMAL( 0, Q[k])
   x[k] = Kalman_correct(xtilde, P_(k,k-1), y[k], theta)
   RETURN x[k]
ENDDEF
\end{verbatim}
Note the intermediate calculation \verb!xtilde!, which is necessary to account for the uncertainty contribution of the state covariance, see Section \ref{sec:batchMC}. 
The sequential Monte Carlo method is then carried out as follows
\begin{enumerate}
  \item Draw $M$ samples $(\bm{x}_0^{(m)},\bm{\theta}^{(m)}), m=1, \ldots , M$, 
	from the PDF $p_{\bm{x}(0), \bm{\theta}}=p_{\bm{x}(0)} p_{\bm{\theta}}$, and set $k=1$
	\label{SMC1}
	\item Draw $M$ samples $(\bm{y}(k)^{(m)}, \bm{z}(k)^{(m)}), m=1, \ldots, M$ from the PDF associated with these input quantities \label{SMC1_1}
\item Evaluate the function \verb!ESTIMATE_Xk! for the drawn samples to obtain $\bm{x}_{k}^{(m)}, m=1, \ldots, M$ \label{SMC2}, and treat the pairs $(\bm{x}^{(m)}_{k},\bm{\theta}^{(m)})$ as samples from
	the PDF $p_{\bm{x}(k), \bm{\theta}}$
	\item Set $k=k+1$ and goto (\ref{SMC1_1})
 \end{enumerate}

When only the mean $\hat{\bm{x}}(k)$ and the covariance matrix $\bm{U}_{\bm{x}(k)}$ and other point-wise statistics are sought, storage of all samples $\{ \bm{x}^{(m)}(k) | m=1,\ldots,M; k=1,\ldots,N \}$ is not necessary. Instead, mean, covariance, credible intervals and other statistics can be calculated from the samples at the current time step and not required sets of samples discarded from computer memory, cf. \cite{Eichstadt2012}.

\section{Particle filtering and related methods}
\label{sec:Particle}
The drawbacks of the extended Kalman filter are a widely discussed topic in the signal processing literature and many alternative approaches have been proposed in the last decades. In particular, methods related to a Monte Carlo sampling for state estimation have been considered by many authors, e.g., \cite{Garcia:2013kq, Douc:2005, Wang:2007ik, Montazeri:2010il}.
In principle, one can translate the probabilistic derivation of state estimation from Section \ref{sec:LKF} to a Bayesian inference for a more general kind of state-space models~\cite{Candy}
\begin{equation}
	p(\bm{x}(k) \vert \bm{y}(k), \bm{x}(k-1)) \propto p(\bm{y}(k) \vert \bm{x}(k)) p(\bm{x}(k) \vert \bm{x}(k-1)) .
\end{equation}
The implementation of the corresponding probability calculus can then be employed by using Markov chain Monte Carlo methods, cf. \cite{Klauenberg:2016dw, Hoffman:2011} or approximation methods \cite{Jiang:2010wg}. In order to take advantage of the sequential character of the filtering process, typically sequential methods are advocated.
For instance, a sequential Markov chain Monte Carlo method can take advantage of the sequential character of the model by reusing information about the Markov chains from the previous time step~\cite{Yang2013}.

However, even then the computational costs are significant due to the high-dimensionality of the time series. Therefore, several approximate sampling methods are proposed in the literature. The most often applied approach is the so-called particle filter, see, e.g. \cite{Garcia:2013kq, Candy, Wang:2007ik} and references therein. It uses a sequential Monte Carlo approach representing the information of the marginal multivariate PDF at the current time instant by a set of weight/sample pairs, called ``particles''. That is, the set $\{ (\bm{x}^{(m)}(k), w^{(m)}(k)) \vert m=1,\ldots,N_s\}$ is considered a discrete approximation of the posterior PDF $p(\bm{x}(k) \vert \bm{y}(k),\bm{x}(k-1))$. The essential tasks of the particle filter are then the calculation of state values $\bm{x}^{(m)}(k)$ and the update of the weights. For the state update the state model can be employed similar as in the Kalman filter prediction step
\begin{equation}\label{123}
	\bm{x}^{(m)}(k) \sim p(f(\bm{x}^{(m)}(k-1),k), \bm{Q}(k))
\end{equation}
with $\bm{Q}(k)$ denoting the covariance matrix of the process noise at time instant $k$. Note that in contrast to the Kalman filter the particle filter does not require the noise to be normally distributed, i.e. the distribution $p()$ in (\ref{123}) does not have to be a Gaussian.
The weight update is based on importance sampling, i.e. sampling from an appropriate density $q(\bm{x}(k)\vert \bm{x}(k-1),y(k))$ and evaluation of $p(\bm{x}(k)\vert \bm{y}(k))$ and $p(\bm{x}(k)\vert \bm{x}(k-1))$
\begin{equation}
	w^{(m)}(k) \propto w^{(m)}(k-1) \frac{p(\bm{y}(k)\vert \bm{x}^{(m)}(k))p(\bm{x}^{(m)}(k)\vert \bm{x}^{(m)}(k-1))}{q(\bm{x}^{(m)}(k)\vert\bm{x}^{(m)}(k-1), \bm{y}(k) )} .
\end{equation}
Thus, the particle filter carries out state estimation without the requirement of linearization or Gaussian PDFs for the noise processes The difference between the particle filter and the GUM Monte Carlo approach is the utilization of the Monte Carlo samples. Whereas the GUM Monte Carlo method aims at propagating samples through a given mathematical model relating the input quantities to the measurand, e.g. the extended Kalman filter, the particle filter has the goal to derive an improved estimate of the states based on the particle weights. Therefore, the particle filter does not correspond to a propagation of uncertainties, but rather to a probabilistic estimation procedure. 
A well-known disadvantage of the particle filter is that the weights deteriorate over time. That is, after a certain number of time steps, a large portion of the samples may have very small weights. Thus, the tails of the distribution are over-represented by the set of particles \cite{Candy}. Many resampling schemes have been proposed as a remedy \cite{Douc:2005}. Such schemes rely on redrawing from the existing set of particles and their subsequent re-weighting. 

Beyond the particle filter, a number of other methods for state estimation for nonlinear state-space models have been proposed, the most famous being the so called unscented Kalman filter \cite{Julier:2004go} and sigma-point filters \cite{VanderMerwe:2004tj}. In \cite{Lefebvre:2004jg} such variants of the Kalman filter are summarized as \emph{linear regression Kalman filters}, because they utilize sampling points for a statistical linear regression of the nonlinear model functions. In the literature, different strategies for choosing appropriate sampling points exist. The accuracy in the approximate evaluation of uncertainties depends on the chosen sampling points \cite{Lefebvre:2004jg}.
Another approach is the representation of the uncertainty associated with the input quantities in terms of intervals and their propagation through the state estimation process \cite{Souto:2009bj}. Although computationally inexpensive, this approach is not compliant with a GUM treatment of uncertainties, because of the strict treatment of uncertainties as bounds rather than probabilistic expressions.

\section{Example}
\label{sec:example}
\begin{figure}[h!]
	\centering 
	\includegraphics[width=\columnwidth]{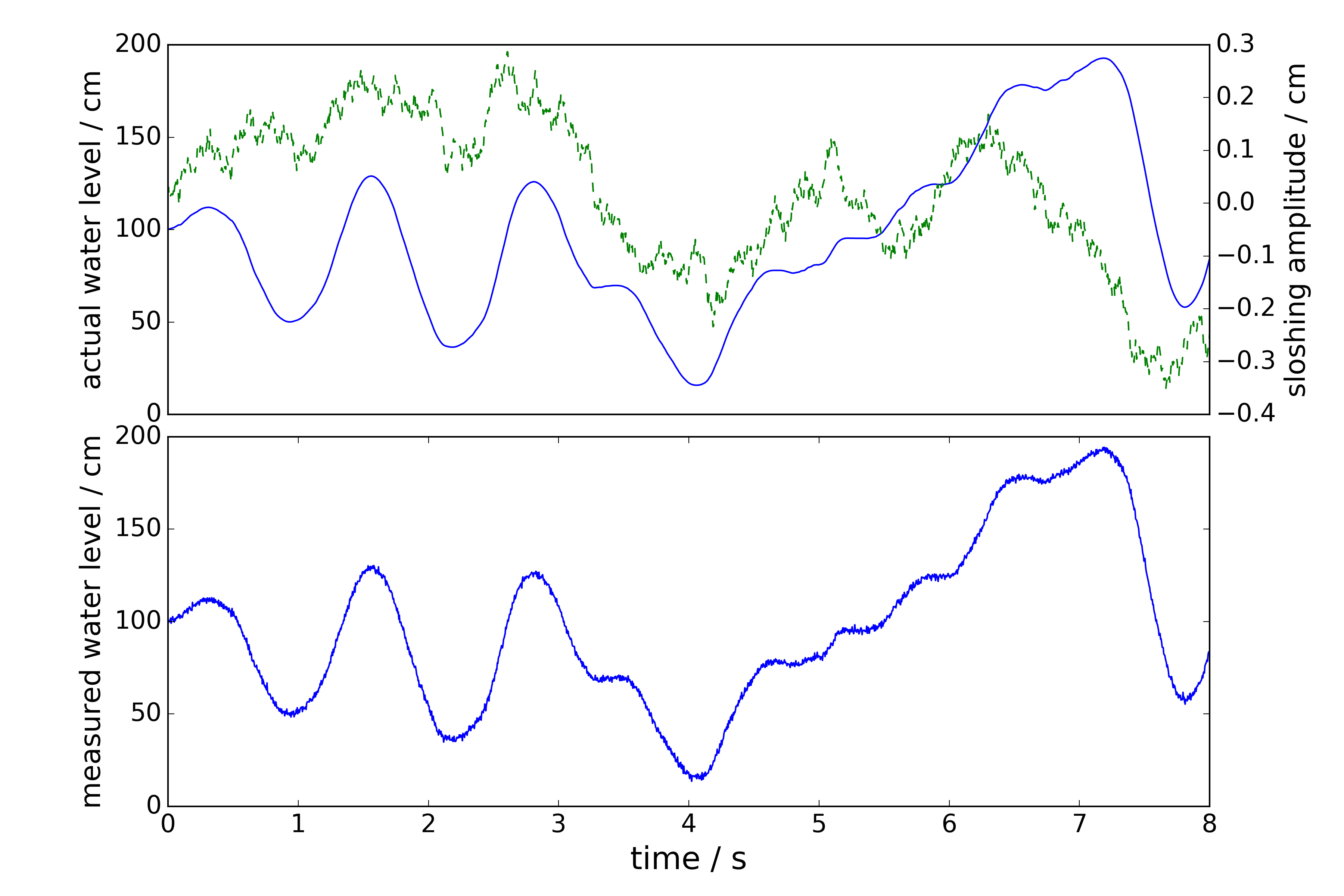}
	\caption{Simulated state-space system \textbf{Top} System states, water level (solid, blue) and sloshing amplitude (dashed, green). \textbf{Bottom} Noisy measurement of water level.}
\label{fig:simulated}
\end{figure}

We illustrate the uncertainty propagation for the Kalman filter using the following example\footnote{We adopted an example from https://www.cs.unc.edu/$\sim$welch/kalman/media/pdf/kftool\_models.pdf}. Consider a water tank where the filling changes in a sinusoidal way, corresponding to a harmonic sloshing. That is, the water level is modeled as
\begin{equation}
	L(t) = L_0 + x_s\sin(2\pi\theta t)
\end{equation}
with initial level $L_0=100$~cm, sloshing amplitude $x_s=0.01$~cm, frequency $\theta=0.8$~Hz and derivative
\begin{equation}
	L'(t) = x_s 2\pi\theta \cos(2\pi\theta t) .
\end{equation}
We consider (noisy) measurements of the fill level and aim at estimating the actual fill level $x_L$ and the sloshing amplitude $x_s$. 
The discrete-time state-space system describing the change of fill level is given by

\begin{align}
\hspace{-10ex}	\left( \begin{array}{c}
x_L(k+1) \\ x_s(k+1)
\end{array}\right)
 =& 
\left(\begin{array}{cc}
1 & 2\pi\theta \cos(2\pi\theta t_k) \\ 0 & 1
\end{array}\right)
\left( \begin{array}{c}
x_L(k) \\ x_s(k)
\end{array}\right)
+
\left( \begin{array}{c}
0 \\ \eta(k)
\end{array}\right) \label{eq:exampleSystem_state}  \\
\hspace{-10ex}y(k) =& \left( 1, 0 \right) \left( \begin{array}{c}
x_L(k) \\ x_s(k)
\end{array}\right)
+ \nu(k) \label{eq:exampleSystem_obs}
\end{align}

where state and measurement noise are modeled as $\eta\sim N(0,\tau^2)$ and $\nu \sim N(0,\sigma^2)$ with standard deviations $\tau = 0.01$ and $\sigma = 1.0$. Figure \ref{fig:simulated} shows the simulated system states over time. 

\subsection{Linear system with known model}

\begin{figure}[h!]
	\centering
	\includegraphics[width=\columnwidth]{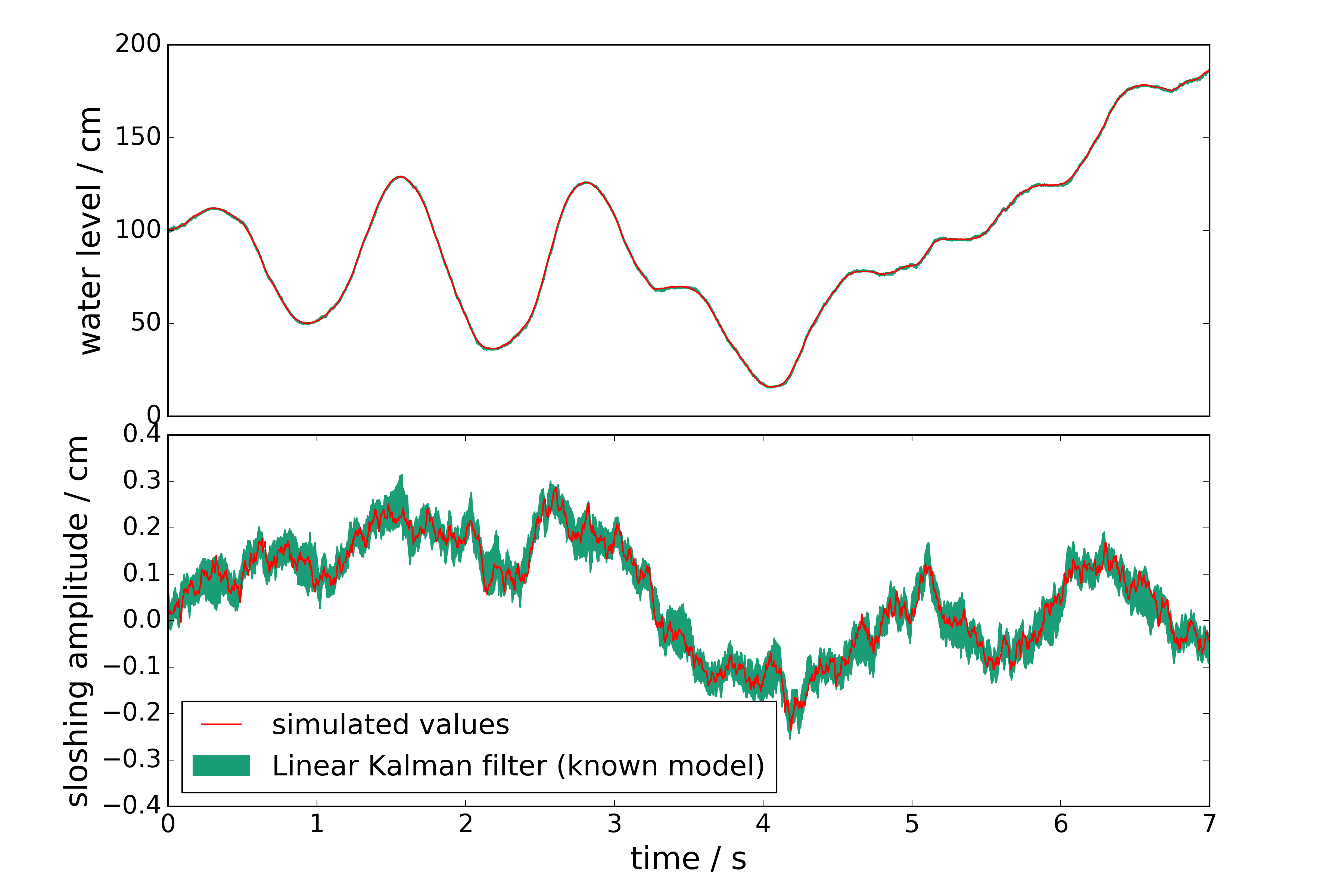}
	\caption{State estimation result of the Kalman filter for the linear system with known model. \textbf{Top} Estimate and point-wise standard uncertainties for the water level. \textbf{Bottom} Estimate and point-wise standard uncertainties for the sloshing amplitude.}
\label{fig:LKF_knownmodel}
\end{figure}

Since the measurement system (\ref{eq:exampleSystem_state})-(\ref{eq:exampleSystem_obs}) is a linear model with normally distributed noise, the linear Kalman filter can be applied for state estimation with
\[ \bm{Q} = \left(\begin{array}{cc}
0 & 0 \\ 0 & \tau^2
\end{array}\right) \qquad \bm{R} = \left( \sigma^2 \right) \qquad \bm{C}=(1,0) \]
and
\[ 
\bm{F}_l(k) = \left(\begin{array}{cc}
1 & 2\pi\theta \cos(2\pi\theta t_k) \\ 0 & 1
\end{array}\right) ~~.
\]
The resulting state estimates are shown in Figure \ref{fig:LKF_knownmodel}. As expected, the Kalman filter estimates correspond very well with the simulated values. For the estimated water level the (k=1) uncertainties are hardly visible at the scale. The estimation would improve even further if smaller measurement noise variances were considered.

\subsection{Linear system with uncertain model}
When the sloshing frequency $\theta$ is not known exactly, the linear Kalman filter is no longer applicable. As described in Section \ref{sec:LKF_uncertain}, two main approaches are considered then, i.e., application of GUM Monte Carlo to the linear Kalman filter or augmentation of the system model (\ref{eq:exampleSystem_state})-(\ref{eq:exampleSystem_obs}) to a nonlinear system.
\subsubsection*{GUM Monte Carlo for linear Kalman filter}
The measurement model for the GUM Monte Carlo method applied to the linear Kalman filter is given as
\begin{equation}
	\bm{x}(k) = \bm{x}_{k,k-1} + \bm{K}(\theta, k)\left( y(k) - \left( 1,0 \right)\bm{x}_{k,k-1} \right)
\label{eq:example_LKFuncertain}
\end{equation}
with 
\[\bm{x}_{k,k-1} = \bm{F}_l(\theta,k)\bm{x}(k-1) + \bm{z}\]
where $\bm{z} \sim N(0,Q)$. The input quantities are then $\bm{x}(0)$, $\theta$ as well as all $\bm{z}(k)$ and $y(k)$. In this example we consider knowledge about $\theta$ to be available in terms of the normal distribution $N(\hat{\theta}, u^2_\theta)$ with $\hat{\theta}$ equal the true value and $u_\theta=0.01~\theta$. The GUM Monte Carlo method is implemented in the sequential way described in Section \ref{sec:MC}. Realizations of $\theta\sim N(\hat{\theta},u^2_{\theta})$ are drawn only once, at the initial time instant, and propagated through the application of the measurement model (\ref{eq:example_LKFuncertain}). The resulting estimates and point-wise standard uncertainties are shown in Figure \ref{fig:LKF_uncertainmodel}.

\subsubsection*{Augmentation to nonlinear model}
When knowledge about the parameter $\theta$ is to be updated using the measurements $y(k)$, the linear system model (\ref{eq:exampleSystem_state})-(\ref{eq:exampleSystem_obs}) is augmented by considering the new state vector $\bm{x}=\left( x_L, x_s, \theta \right)^{\mathsf{T}}$:
\begin{align}
\hspace{-10ex}	\left( \begin{array}{c}
x_L(k+1) \\ x_s(k+1) \\ \theta(k+1)
\end{array}\right)
 =& 
\left(\begin{array}{c}
x_L(k)+ x_s(k)2\pi\theta(k) \cos(2\pi\theta(k) t_k) \\ x_s(k) \\ \theta(k)
\end{array}\right)
+
\bm{\eta}(k) \\
\hspace{-10ex}y(k) =& \left( 1, 0, 0\right) \left( \begin{array}{c}
x_L(k) \\ x_s(k) \\ \theta(k)
\end{array}\right)
+ \nu(k) .
\end{align}
This model is nonlinear, and hence the linear Kalman filter cannot be applied. Therefore, the extended Kalman filter is considered
with initial state estimate and its associate uncertainty given as
\begin{equation}
	\bm{x}_0 = \left( \begin{array}{c}
L_0, 0.01, \hat{\theta} 
\end{array}\right)
\qquad
\bm{P}_0 = \left(\begin{array}{ccc}
0 & 0 & 0 \\
0 & \tau^2 & 0 \\
0 & 0 & u_\theta^2
\end{array}\right) ~~.
\end{equation}

The extended Kalman filter state estimate is then calculated as
\begin{equation}
	\bm{x}(k) = \bm{x}_{k,k-1} + \bm{K}(k) \left( y(k) - \left( 1,0,0 \right)\bm{x}_{k,k-1} \right)
\label{eq:example_EKF}
\end{equation}
with
\[
\bm{x}_{k,k-1} = f(\bm{x}(k-1), k) + \bm{z}
\]
where $\bm{z}\sim N(0,\bm{\tilde{Q}})$. The state augmentation ideally yields $\bm{\tilde{Q}}=diag((0, \tau^2, 0))$, modelling $\theta$ as completely time-invariant. In our example, however, the extended Kalman filter tended to perform significantly better when introducing an additional process noise
with variance $\alpha^2 << u_\theta^2$, i.e., using $\bm{\tilde{Q}}=diag((0,\tau^2, \alpha^2))$. That the process noise is a crucial ``tuning'' parameter of the extended Kalman filter and related methods is well-known and different strategies to choosing optimal process noise covariances are available in the literature \cite{Bavdekar:2011it, Chatzi:2009jp}.

\begin{figure}[h!]	
	\centering
	\includegraphics[width=\columnwidth]{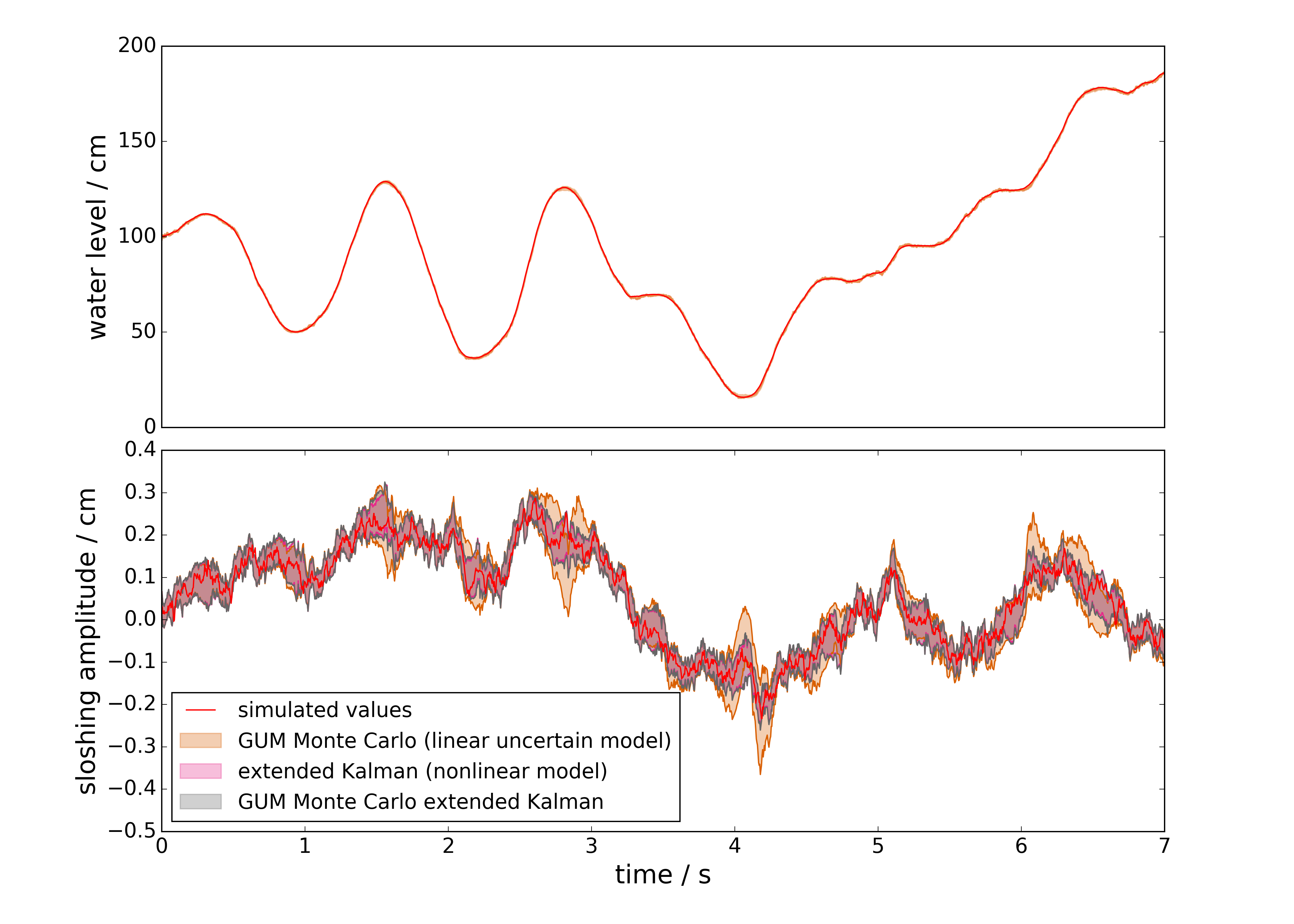}
	\caption{State estimation result for the linear system with model uncertainty, and for the corresponding nonlinear model. Shown are the estimate and point-wise standard uncertainties for the water level (top) and the sloshing amplitude (bottom).}
	\label{fig:LKF_uncertainmodel}
\end{figure}
 
As described in Section \ref{sec:LKF_uncertain}, in order to render equation (\ref{eq:example_EKF}) a GUM-compliant measurement model, the parameter $\theta$ has to be considered as time dependent. That is, $\theta$ is an input quantity only for the initial time step. With this adaptation, the GUM Monte Carlo method can be applied to the measurement model (\ref{eq:example_EKF}) with input quantities being 
$\bm{x}(0)$ and all $\bm{z}(k), y(k)$.

\begin{figure}[h!]
	\centering
	\includegraphics[width=\columnwidth]{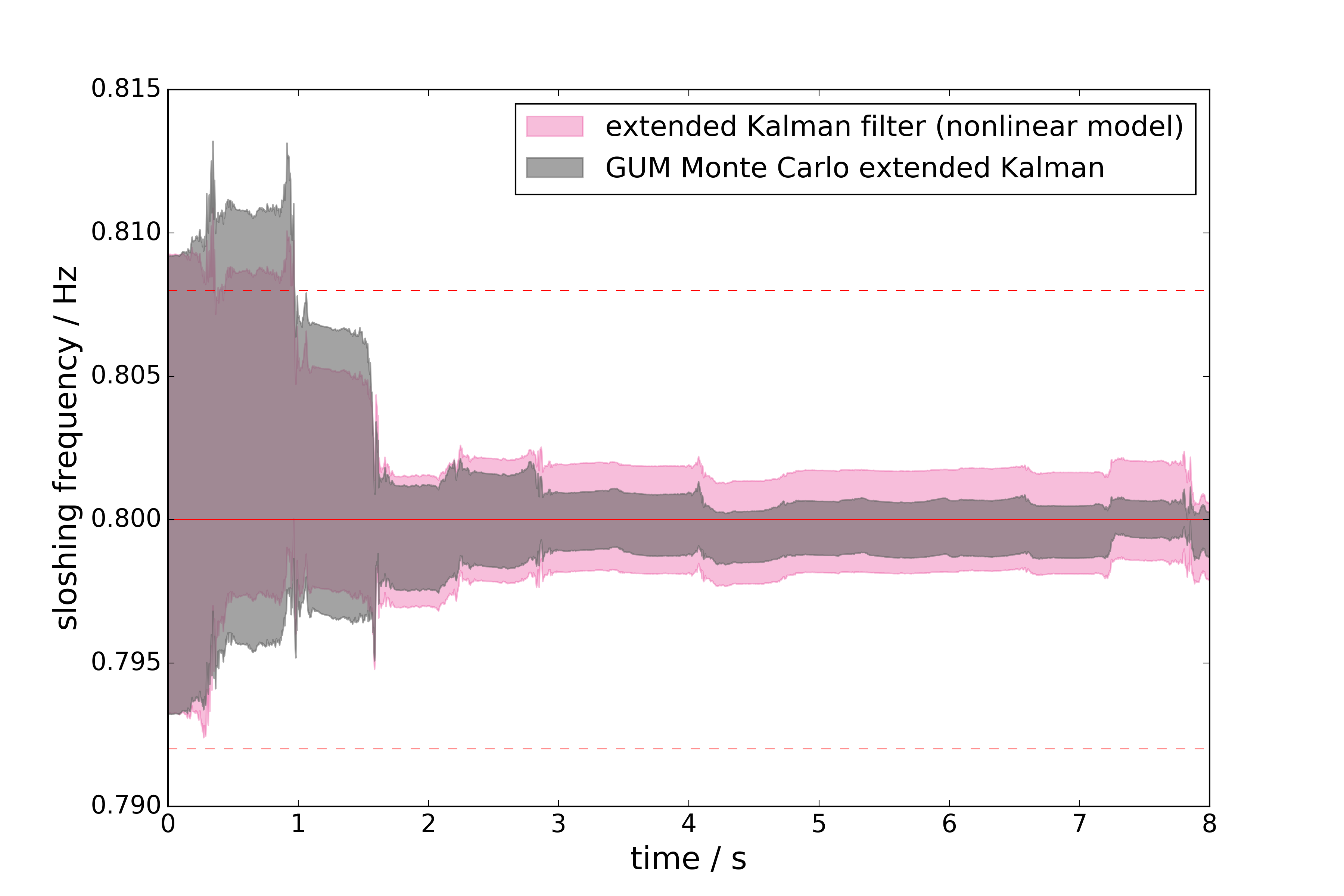}
	\caption{Sloshing frequency estimated by the extended Kalman filter with and without uncertainty evaluation carried out by the GUM Monte Carlo method.}
	\label{fig:EKFvsMCEKF}
\end{figure}

Figure~\ref{fig:LKF_uncertainmodel} shows the estimate and associated point-wise standard uncertainties for the original state variables $x_L$ and $x_s$, for the three approaches to the treatment of model parameter uncertainty, i.e., GUM Monte Carlo for the linear model, extended Kalman filter and GUM Monte Carlo for the extended Kalman filter. For the Monte Carlo methods we carried out $100$~$000$ trials.

The estimated water level $x_L$ is almost identical for all three approaches, whereas the uncertainties of the estimated sloshing amplitudes are slightly larger for the GUM Monte Carlo method applied to the linear model. The reason is that this approach does not alter the model parameter estimate nor its associated uncertainty. In contrast, the extended Kalman filter, and consequently the GUM Monte Carlo for the extended Kalman filter, calculate a new value of $\theta$ and its associated uncertainty in every time step. Provided that the extended Kalman filter converges, an improved estimate of the model parameter can potentially improve the overall estimation quality.

The estimates and standard uncertainties for the uncertain model parameter are shown in Figure \ref{fig:EKFvsMCEKF}, showing that the uncertainty associated with the sloshing frequency $\theta$ is decreased by the extended Kalman filter based on the sequential re-estimation of this parameter. That is, after a small number of time steps the extended Kalman filter shows an increased confidence in the estimated value of $\theta$. The uncertainties calculated by the GUM Monte Carlo for the measurement model (\ref{eq:example_EKF}) resembles that of the extended Kalman filter, but are significantly smaller after a certain number of time steps.

\subsubsection*{Approximate Bayesian inference}
In general, the extended Kalman filter suffers from linearization errors which cannot be mitigated by an application of GUM Monte Carlo for the corresponding uncertainty evaluation, unless this error can be quantified and included in the uncertainty budget. Therefore, we compared the extended Kalman filter result with that obtained by a particle filter, implemented by carrying out the following calculations in each time step, where $m$ denotes the index of the particle~\cite{Arulampalam:2002hg}.
\begin{enumerate}
 \item Calculate a new estimate of the state variable as \[\bm{x}^{(m)}(k) = f(\bm{x}^{(m)}(k-1),k) + \bm{\eta}(k) \] with $\bm{\eta}(k)\sim N(0,\tilde{\bm{Q}}(k))$
 \item With $\hat{y}(k)$ the measurement at time instant $k$, evaluate the likelihood $l(\bm{x}^{(m)}(k);\hat{y}(k))$.
 \item Update the weights as \[w^{(m)}(k) = w^{(m)}(k-1)\cdot l(\bm{x}^{(m)}(k) ; \hat{y}(k))\]
 \item Calculate the approximate effective sample size as \[\hat{N}_{eff} = \left( \sum_{m} (w^{(m)}(k))^2 \right)^{-1} \]
 \item Apply multinomial resampling if $\hat{N}_{eff} < \gamma N_s$ with $N_s$ the actual sample size and $\gamma\in (0,1]$ some chosen tolerance factor
\end{enumerate} 

\begin{figure}[h!]
	\centering
	\includegraphics[width=\columnwidth]{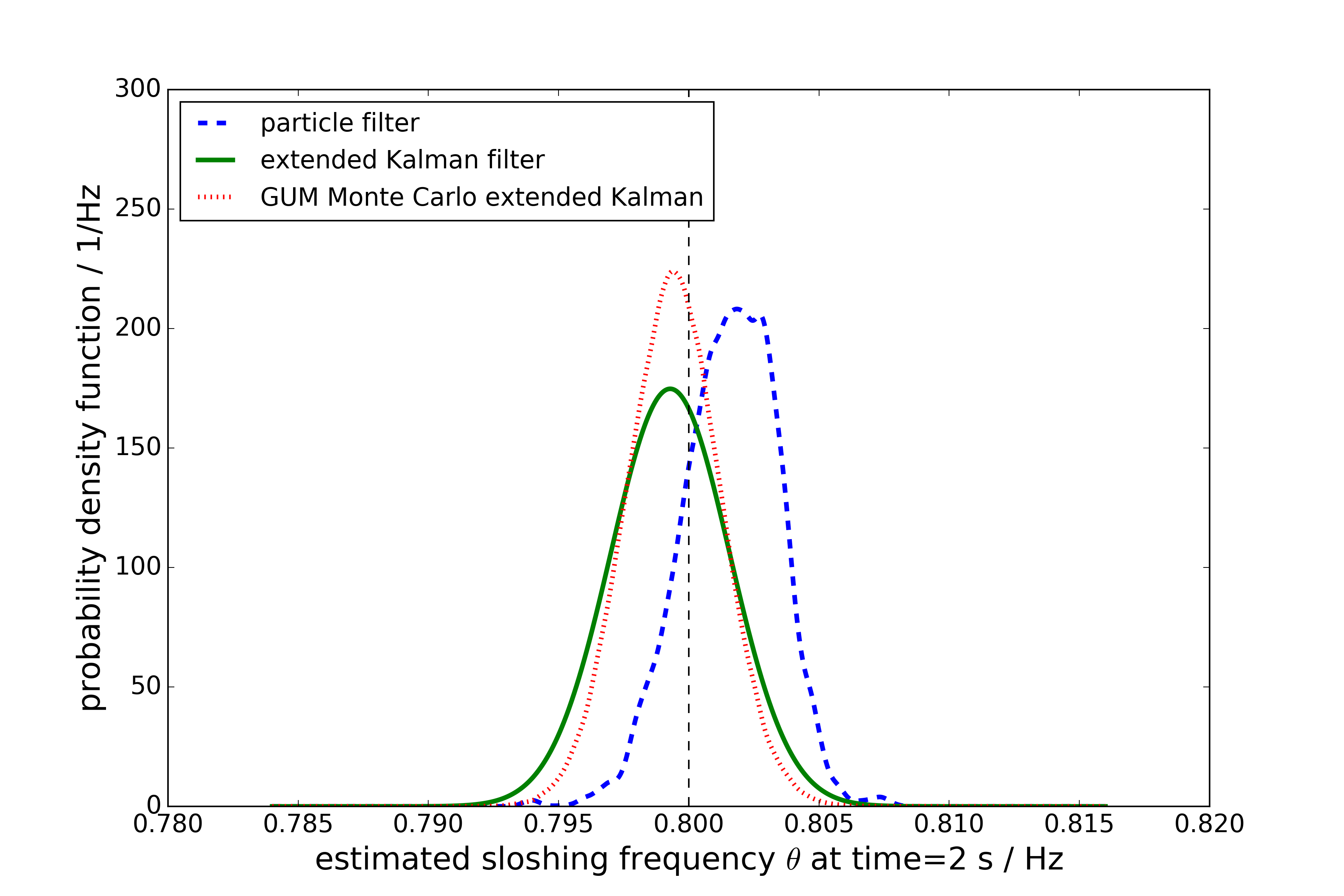}\\
	\includegraphics[width=\columnwidth]{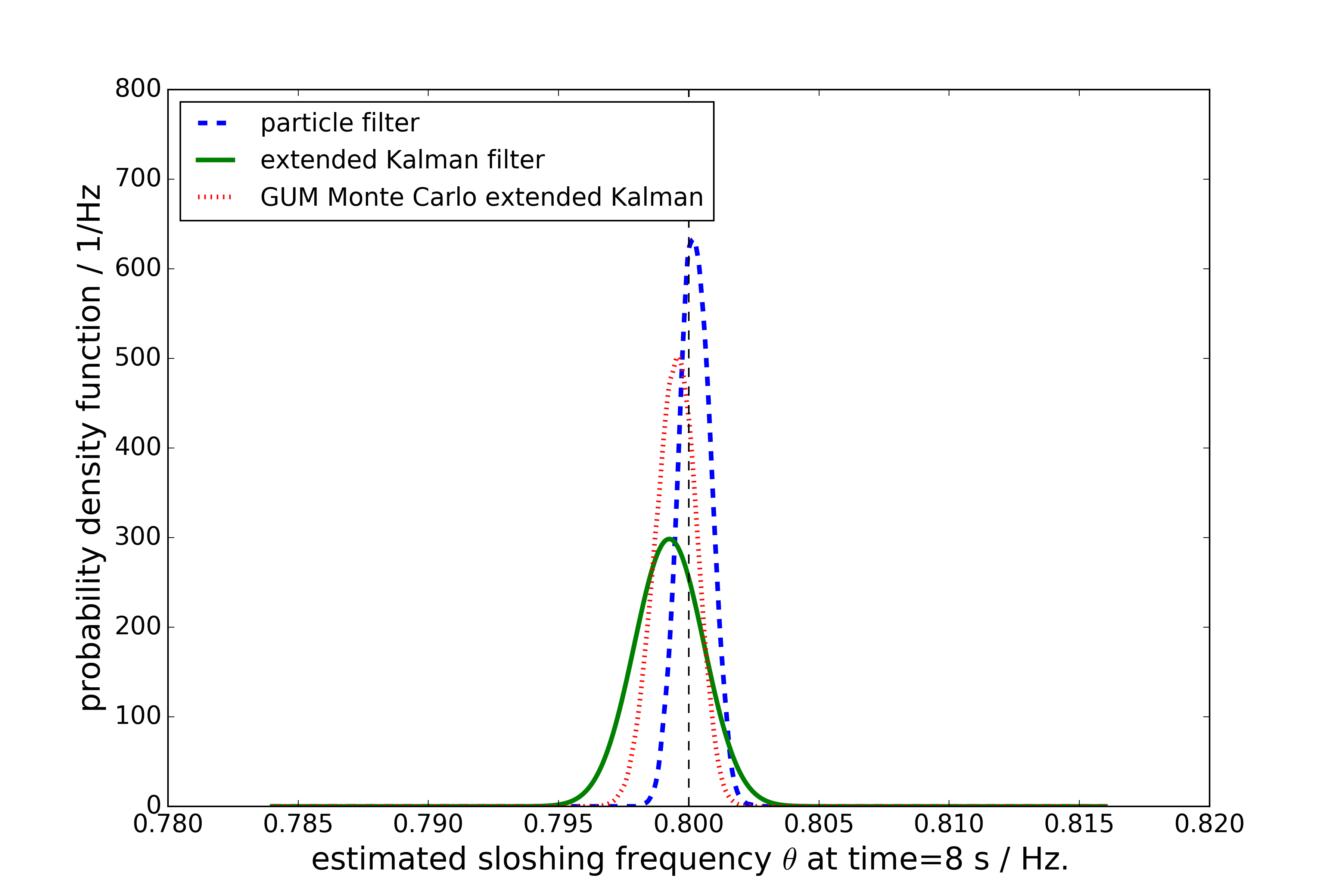}
	\label{fig:PFresults}
	\caption{Result of the particle filter with $N_s=10~000$ samples (dashed) and that of the extended Kalman filter (solid) \textbf{Top} at time instant $t=2 s$ and \textbf{Bottom} at time instant $t=8 s$.}
	\label{fig:PFresults}
\end{figure}

The plots shown in Figure \ref{fig:PFresults} illustrate the change in the marginal PDF associated with the sloshing frequency $\theta$ as determined by the particle filter with $N_s=100~000$ particles and resampling tolerance factor $\gamma=0.9$. As for the extended Kalman filter, the estimation is initialized with $\bm{P}_\theta(0)=u_\theta^2$ as initial variance, which decreased during the estimation process over time. 
Figure \ref{fig:PFresults} shows that at time equal $8$~s the PDF calculated by the particle filter is centered around the true value of $\theta$, whereas the PDF obtained from the GUM Monte Carlo method for measurement model (\ref{eq:example_EKF}) is slightly shifted. Also shown in Figure~\ref{fig:PFresults} is the normal distribution obtained from the estimate and covariance matrix calculated by the extended Kalman filter. This PDF is also not centered around the true value at this point and has a larger spread than the PDFs obtained from the particle filter and the GUM Monte Carlo method.
The difference between the PDFs obtained at time equal $8$~s indicate a faster convergence rate of the particle filter in the example considered, with computational costs being comparable to those of the GUM Monte Carlo method. 

\section{Conclusions}
\label{sec:conclusions}
The Kalman filter and related methods for state estimation are successfully employed in a wide range of areas. The Kalman filter sequentially produces an estimate of the state vector and characterizes its uncertainty by a covariance matrix that is updated during the sequential estimation.
An implementation of the Kalman filter for metrological applications requires a reliable evaluation of the uncertainty associated with the obtained state estimate. Since the GUM \cite{GUM} and its Supplements provide the de facto standard for uncertainty evaluation in metrology, the question arises whether the covariance matrix produced by the Kalman filter is a GUM-compliant uncertainty characterization. 
Recent supplements of the GUM \cite{GUMS1, GUMS2} advocate the use of state of knowledge distributions for the characterization of the uncertainty about a measurand. Therefore, the equivalence or in-equivalence of a GUM uncertainty evaluation for Kalman state estimation and a Bayesian inference is of interest as well. We have addressed these issues and compared the different approaches for linear as well as nonlinear dynamic systems. In addition, the situation of not precisely known systems has been considered which is particularly important for dynamic metrology \cite{Eichstadt2012Diss}.\\

For linear, exactly known dynamic systems, the Kalman filter provides optimal estimates in the sense of root mean square error \cite{Kalman}. For this case, we showed that the covariance matrix produced by the Kalman filter is GUM-compliant. Furthermore, the results are also equivalent to a Bayesian inference. However,
this general equivalence breaks down when the system becomes nonlinear. These findings are similar to the situation of regression models. There, the GUM Monte Carlo method applied to a least-squares model is equivalent to a Bayesian inference provided that the regression model is linear in its parameters (and that a particular noninformative prior is employed); for nonlinear regression models also different results are obtained in general \cite{Elster:2011jo}.\\

When a linear system is no longer known exactly but depends on further parameters, the GUM Monte Carlo method can be applied in connection with a GUM model defined by the linear Kalman filter estimation, accounting for the incomplete knowledge of the additional parameters. The complete model then, however, is in general nonlinear. Application of the GUM Monte Carlo method then provides a GUM-compliant uncertainty analysis that does not suffer from linearization errors. An appropriate, sequential algorithm for the implementation of such a GUM Monte Carlo method was proposed which can be applied for real-time applications.\\

For nonlinear, exactly known dynamic systems the results obtained by the extended Kalman filter correspond to those obtained by the GUM propagation of covariance matrices upon linearization. They are, however, no longer equivalent to the application of GUM Monte Carlo. The latter approach can improve the uncertainty results obtained by the extended Kalman filter to some extent. However, as the GUM model is based on a linearization of the dynamics (through the extended Kalman filter estimation), a GUM Monte Carlo approach cannot mitigate shortcomings due to that linearization. An alternative then is the application of a Bayesian inference which, however, can be expensive to implement. Nevertheless, an approximate Bayesian inference such as the considered particle filtering approach may provide an attractive alternative in this case. Similar conclusions hold when the nonlinear dynamic system is not known exactly.\\

In summary: For linear, exactly known systems application of the Kalman filter can be recommended as the method of choice for state estimation, providing a GUM-compliant uncertainty characterization which also is equivalent to a Bayesian inference. For linear, not exactly known systems we recommend to use the proposed sequential GUM Monte Carlo method based on the linear Kalman filter estimation. In the case of nonlinear, not exactly known dynamic systems both the extended Kalman filter as well as the application of a GUM uncertainty analysis based on the extended Kalman filter show shortcomings, and an approximate Bayesian method such as particle filtering may yield more adequate results then.

\section*{Acknowledgement}
This work is part of the European Metrology Research Program (EMRP) Joint Research Project ENG63. The EMRP is jointly funded by the EMRP participating countries within EURAMET and the European Union.

\bibliographystyle{unsrt}
\bibliography{literature}

\end{spacing}

\end{document}